\documentclass[twoside,12pt]{amsart}
\usepackage{a4wide}
\usepackage{amsmath}
\usepackage{amsfonts}
\usepackage{amssymb}
\usepackage{times} 
\usepackage{pstricks}
\usepackage{pst-node}

\psset{framearc=0.0,framesep=0.0truecm,nodesep=3pt,unit=0.75truecm,%
arrowsize=4pt 2,arrowinset=0.4}
\def\pstlw{0.8pt}
\newcounter{mycnt}
\def\themycnt{\thesection.\arabic{mycnt}}
\def\mybenv#1{\refstepcounter{mycnt}%
       \vskip 3pt\noindent{\bf \themycnt~~#1}:~}
\def\myeenv{\hfill\rule{1ex}{1ex}\vskip 3pt}
\def\qed{\hfill$\Box$}
\def\nn{\nonumber \\}
\def\Id{\text{I\!d}}

\def\openC{\mathbb{C}}
\def\openN{\mathbb{N}}
\def\openZ{\mathbb{Z}}
\def\openP{\mathbb{P}}
\def\openQ{\mathbb{Q}}
\def\antip{{\sf S}}

\def\DP{\Delta^{+}}
\def\DPU{\underline{\Delta}^{+}}
\def\DM{\Delta^{\cdot}}
\def\DMU{\underline{\Delta}^{\cdot}}
\def\SP{\textsf{S}^{+}}
\def\SM{\textsf{S}^{\cdot}}
\def\SMU{\underline{\textsf{S}}^{\cdot}}
\def\conv{\star}
\def\lcm{\mathrm{lcm}}
\def\Conv{\mathrm{Conv}}
\def\la{\langle}
\def\ra{\rangle}
\def\pprime{\prime\prime}
\def\ds{\oplus}
\def\!{\kern -0.15ex}
\def\Hom{\text{hom}}

\begin{document}

\title{Renormalization : A number theoretical model} 

\author{Bertfried Fauser}
\address{Max Planck Institute for Mathematics in the Sciences, 
Inselstrasse 22-26, D-04103 Leipzig, Germany}
\email{fauser@mis.mpg.de}

\subjclass[2000]{Primary 16W30; Secondary 30B50; 11A15; 81T15; 81T16}

\keywords{Hopf algebra, renormalization, Dirichlet series,
Dirichlet convolution, multiplicative arithmetic functions}

\date{January 25, 2006}


\begin{abstract}
We analyse the Dirichlet convolution ring of arithmetic number theoretic
functions. It turns out to \textit{fail} to be a Hopf algebra on the diagonal,
due to the lack of complete multiplicativity of the product and coproduct.
A related Hopf algebra can be established, which however overcounts the
diagonal. We argue that the mechanism of renormalization in quantum field
theory is modelled after the same principle. Singularities hence arise as a
(now continuously indexed) overcounting on the diagonals. Renormalization is
given by the map from the auxiliary Hopf algebra to the weaker multiplicative
structure, called Hopf gebra, rescaling the diagonals.
\end{abstract}

\maketitle

{\small\tableofcontents}

\section{Dirichlet convolution ring of arithmetic functions}

\subsection{Definitions}

In this section we recall a few well know facts about formal 
Dirichlet series and the associated convolution ring of Dirichlet 
functions \cite{apostol:1979a,bruedern:1995a}. An arithmetic function
is a map $f : \openN\rightarrow \openC$. Equivalently we can consider
integer indexed sequences of complex numbers. It is convenient to
introduce formal generating functions to encode this information in
a more compact form
\begin{align}
f(s) &:=\sum_{n\ge 1} \frac{f(n)}{n^{s}} \nn
  s  &= \sigma + i\,t \in \openC
\end{align}
where the formal complex parameter is traditionally written as $s$. No 
confusion should arise between the series elements $f(n)$ and the generating
function $f(s)$ formally denoted in the same way.

A ring structure is imposed in the obvious manner:
\mybenv{Definition}
The Dirichlet convolution ring of arithmetic functions is defined
on the set of arithmetic functions as
\begin{align}
(f+g)(s) &:= \sum_{n\ge 1} \frac{f(n)+g(n)}{n^s} \nn
(f\conv g)(s) &:= \sum_{n\ge 1} \sum_{d\vert n}
\frac{f(d)\cdot g(n/d)}{n^s} 
\end{align}
Where $\sum_{d\vert n}$ is the sum over all divisors $d$ of $n$. 
\myeenv
The component wise addition imposes a module structure on the arithmetic
functions, and the convolution product is actually the point wise product
of the generating functions $f(s)\cdot g(s)$ as is easily seen. Furthermore
the product is commutative, associative, and unital with unit $u(s):=
\sum_{n\ge 1} \delta_{n,1}n^{-s}$ where $\delta_{n,1}$ is the Kronecker
delta symbol.

If $f(1)\not=0$ then a unique inverse Dirichlet generating function exists
w.r.t. the convolution product 
\begin{align}
f\conv f^{-1} 
&=u=f^{-1}\conv f \nn
n=1:\quad f^{-1}(1)&=1/f(1)\nn
n>1:\quad f^{-1}(n)&=
 \frac{1}{f^{-1}(1)}\sum_{{d\mid n}\atop{d<n}} f^{-1}(\frac{n}{d})f(d)
\end{align}
The invertible arithmetic functions form a group:
\begin{align}
f\conv u&=f=u\conv f \nn
(f\conv g)^{-1} &= g^{-1}\conv f^{-1}
\end{align}
due to the associativity of the convolution.

\subsection{Multiplicativity versus complete multiplicativity}

Two integers $n,m$ are called relatively prime if their greatest common 
divisor $\gcd(n,m)=(n,m)$ is $1$ hence if they have no prime
factor in common. Many important number theoretical functions
enjoy a weakened homomorphism property, called multiplicativity.
\mybenv{Definition}
An arithmetic function $f$ is called \textbf{complete multiplicative} if
\begin{align}
f(n\cdot m)&=f(n)\cdot f(m)\hskip 1truecm \forall n,m
\end{align}
An arithmetic function $f$ is called \textbf{multiplicative} if
\begin{align}
f(n\cdot m)&=f(n)\cdot f(m)\hskip 1truecm 
  \forall n,m \text{~with~} (n,m)=1
\end{align}
\myeenv
Hence multiplicative functions fail in general to be homomorphisms of the
multiplicative structure of the natural numbers iff the product has a
nontrivial common prime number content $(n,m)=k$, such that
$n=k\cdot n^\prime$ and $m=k\cdot m^\prime$. We may call $k$ the overlapping
or meet part of $n,m$. Actually $\gcd$ and $\lcm$ form a distributive lattice
on the integers.

\subsection{Examples}

We give some examples of arithmetic functions, among them multiplicative,
complete multiplicative and non-multiplicative ones, which all play important
roles in number theory.

Let $n=\prod p_i^{r_i}$, $m=\prod p_i^{s_i}$ and define $\nu$ to be the 
function $\nu(n)=2^{\sum r_i}$. One has
\begin{align}
\nu(n\cdot m) 
&= \nu(\prod p_i^{r_i+s_i}) 
 = 2^{\sum (r_i+s_i)} = 2^{(\sum r_i)+(\sum s_i)} \nn
&= 2^{\sum r_i} \, 2^{\sum s_i} = \nu(n)\, \nu(m)
\end{align}
Hence $\nu$ is a homomorphism or complete multiplicative function.

The M\"obius function is defined as
\begin{align}
\mu(n)&=\left\{
\begin{array}{cl}
1 & n=1 \\
0 & n \text{~~contains a square} \\
(-1)^k & n=\prod^k_{i=1} p_i, \text{~~$k$~distinct~primes}
\end{array}
\right.
\end{align}
The sequence of integer values of the M\"obius function is a random-looking
list of ${\pm 1,0}$ entries:
\begin{align}\label{moebiusTab}
\begin{array}{c|rrrrrrrrr}
     n & 1 & 2 & 3 & 4 & 5 & 6 & 7 & 8 & \ldots \\ \hline
\mu(n) & 1 &-1 &-1 & 0 &-1 & 1 &-1 & 0 & \ldots 
\end{array}
\end{align}
Another interesting arithmetic function is the Euler totient function, which
counts the number of relative prime numbers $d$ having $(d,n)=1$ smaller than 
$n$. Using $\#$ for cardinality it reads
$\phi = \#\{d\in \openN;\,\, d< n,\,\, (d,n)=1\}$:
\begin{align}
\begin{array}{c|ccccccccc}
n       & 1 & 2 & 3 & 4 & 5 & 6 & 7 & 8 & \ldots \\ \hline
\phi(n) & 1 & 1 & 2 & 2 & 4 & 2 & 6 & 4 & \ldots 
\end{array}
\end{align}
Introducing the arithmetic function $N(n)=n$, $\forall n$ one finds 
$\phi(n)=(\mu\conv N)(n) = n\prod_{p\mid n}(1-\frac{1}{p})$. The M\"obius and
Euler totient functions are multiplicative, but not complete multiplicative.

A further example of a non multiplicative function is the von Mangoldt 
function:
\begin{align}
\Lambda(n) &= 
\left\{
\begin{array}{cl}
\log p & \text{if $n=p^m$,~~ $m \ge 1$,~~} p \text{~~a prime} \\
0 & \text{otherwise (including $1$)}
\end{array}
\right.
\end{align}
Tabulated this reads
\begin{align}
\begin{array}{c|ccccccccc}
         n & 1 & 2 & 3 & 4 & 5 & 6 & 7 & 8 & \ldots \\ \hline
\Lambda(n) & 0 & \log 2 & \log 3 & \log 2 & \log 5 & 
                   0 & \log 7 & \log 2 & \ldots 
\end{array}
\end{align}
The importance of the von Mangoldt function stems from the fact that it
encodes the derivation with respect to the formal parameter $s$ of a
Dirichlet generating function in terms of the convolution product. We
use $\sum_{d\vert n} \Lambda(d)=\log n$ to show this
\begin{align}
\frac{\partial}{\partial s}f(s)
&=\sum_{n\ge 1} f(n) \frac{\partial}{\partial s}n^{-s} 
 =\sum_{n\ge 1} f(n)(-\log n)n^{-s}
\end{align}
In particular one obtains for the Riemann zeta function $\zeta^{-1}=\mu$,
$\zeta(n)=1$ $\forall n$ the formula
\begin{align}
-\frac{\zeta(s)^ \prime}{\zeta(s)} 
&= -\frac{\partial}{\partial s} \log \zeta(s)
 = \Lambda(s) 
\end{align}
The von Mangoldt function appears in the Selberg formula
\cite{selberg:1949a}, which allows one to embark on an `elementary', 
that is nonanalytic, proof of the prime number theorem.
\myeenv

\section{Products and coproducts related to Dirichlet convolution}
\setcounter{equation}{0}

In previous work, we studied extensively the Dirichlet Hopf algebra
of arithmetics \cite{fauser:jarvis:2006a}. We extract from that work
the two coproducts needed for the present purpose. We dualize the 
(semi)ring structure\footnote{We will later on always complete the natural
numbers \`a la Grothendieck to a group, the integers, hence `ring' will be 
\textit{a posteriori} justified.} of the natural numbers $(\openN,+,\cdot)$
using the Kronecker duality written as a scalar product
$\la\ \mid\ \ra : \openN \times \openN \rightarrow \openZ_2$,
$\la n\mid m\ra =\delta_{n,m}$.
\mybenv{Definition} 
The coproduct of addition is defined as
\begin{align}
\DP(n) 
&:=\sum_{n_1+n_2=n}n_1 \ds n_2 \nn
 &= n_{(1)} \ds n_{(2)}
\end{align}
and the coproduct of multiplication is defined as
\begin{align}
\DM(n) 
&:=\sum_{n_1\cdot n_2=n}n_1 \times n_2=\sum_{d\mid n} d\times \frac{n}{d}\nn
 &=n_{[1]}\times n_{[2]}
\end{align}
We introduced Sweedler indices and the Brouder-Schmitt convention
\cite{brouder:schmitt:2002a} to denote Sweedler indices of different coproducts
by different parentheses.
\myeenv
Our coproducts allow us to define convolution algebras on the (set of
coefficients of) arithmetic functions $f : \openN\rightarrow \openC$ together
with the product $\mu$ in $\openC$. If the codomain of such a function is in
the nonnegative integers, it is an endomorphism of $\openC$ under a suitable
identification of $\openN$ in $\openC$. The process of extending the convolution
of (set) endomorphisms on $\openN$ to (set) homomorphisms 
$\Hom(\openN,\openC)$ is subtle, since ring extensions have to be considered.
In our formal treatment we do not care about this.
\mybenv{Definition} A convolution algebra $\Conv(\mu,\Delta)$ is defined on
homomorphisms $f,g\in\Hom(\openN,\openC)$ as
\begin{align}
(f+g)(n) &= `+'(f\times g)\DP(n) \,=\, f(n)+g(n)\nn
(f\conv g)(n) &= \mu(f\times g)\Delta(n) 
\,=\, \sum_{d\mid n} f(n)\cdot g(\frac{n}{d})
\end{align}
for the product $\mu$ and coproduct $\DP$ respectively $\DM$.
\myeenv
It is easy to show the following
\mybenv{Proposition}
$\Conv(+,\DP)$ is biassociative, biunital, bicommutative with antipode
$\SP : \openN\rightarrow \openZ$ given by $\SP(n)=-n$
\myeenv
Note that the antipode is $\openZ$ valued forcing us to extend the codomain
of the homomorphisms at least to $\openZ$. We introduce the Hadamard
product $. : \Hom(\openN,\openC)\times \Hom(\openN,\openC)\rightarrow
\Hom(\openN,\openC)$, that is the coefficient wise product of Dirichlet series, 
as $(f.g)(s) =\sum_{n\ge 1}f(n)\cdot g(n) n^{-1}$ to be able to state the
\mybenv{Proposition}
$\Conv(\cdot,\DM)$ is biassociative, biunital, bicommutative with antipode
$\SM : \openN\rightarrow \openZ$ given by $\SM(n)=(N.\mu)(n) = 
n\cdot\mu(n)$ or alternatively written as generating function
$\SM(s) = \mu(s-1)$.
\myeenv

While the first statements are almost trivial, the antipode can be derived
as a group inverse using a recursion argument. Tabulated it reads
\begin{align}
\begin{array}{c|rrrrrrrrr}
     n & 1 & 2 & 3 & 4 & 5 & 6 & 7 & 8 & \ldots \\ \hline
\SM(n) & 1 &-2 &-3 & 0 &-5 & 6 &-7 & 0 & \ldots 
\end{array}
\end{align}
which should be compared with the table (\ref{moebiusTab}).

The coproduct of multiplication models exactly the split of arguments
in the Dirichlet convolution. In this case the Hopf algebraic version
acts directly on the elements of the series representation of the
arithmetic functions. The remarkable fact is, that this coproduct can be
obtained from an almost trivial dualization of multiplication of integers.
The coproduct of addition will come into play later. We want to make this
duality explicit, using the Kronecker pairing $\la n\mid m\ra=\delta_{n,m}$
\begin{align}
\la n + m\mid k\ra 
&= \la n\ds m\mid \DP(k)\ra
 = \la n\mid k_{(1)}\ra\la m\mid k_{(2)}\ra \nn
\Leftrightarrow\quad \DP(k)
&=\sum_{k_1+k_2=k} k_1\times k_2 =k_{(1)} \ds k_{(2)}
\end{align}
And for the coproduct of multiplication one has
\begin{align}
\la n\cdot m\mid k\ra 
&= \la n\times m\mid \DM(k)\ra
 = \la n\mid k_{[1]}\ra\la m\mid k_{[2]}\ra \nn
\Leftrightarrow\quad \DM(k)
&=\sum_{k_1\cdot k_2=k} k_1\times k_2
 =\sum_{d\vert n} d\times \frac{n}{d}
 =k_{[1]} \times k_{[2]}
\end{align}
We close this discussion by exhibiting the primitive elements with respect
to coaddition and comultiplication. A $(1,1)$-primitive element (or simply
primitive element) $p$ is defined satisfying the relation
$\Delta(p)=p\otimes 1 +1 \otimes p$. Using the particular monoidal structure,
i.e. direct sums for addition, cartesian product for multiplication, and the
respective units for addition and multiplication, we find as an easy 
consequence of the definitions:
\mybenv{Corollary}
With respect to the coproduct of addition $\DP$, $1$ is the only
primitive element and $\openN$ is additively generated by $1$.
\myeenv
\mybenv{Corollary}
With respect to the coproduct of multiplication$\DM$, $\{p_i\}$ 
the set of all prime numbers represents all primitive elements and 
$\openN$ is multiplicatively generated by these primes.
\myeenv
This poses the opportunity to introduce two gradings on $\openN$ turning the
integers into a graded set. First by setting $\openN = \ds_{n\in\openN} 
{\bf 1}^n$ where every number represents its own grade. Addition is a
graded map (binary `product') under this grading. Now let $\openP$ be the
set of all prime numbers and $\openP^k$ the set of all integers having
exactly $k$ prime factors (including multiplicities). Let $\openP^0=1$.
The grading suggested by the multiplicative structures is defined as:
\begin{align}
\openN = \oplus_{i\ge0} \openP^i
\end{align}
This regrouping will have a great influence on how densities or the
asymptotic behaviour of Dirichlet arithmetic functions have to be
considered, see appendix \ref{Petermann}. For a detailed discussion 
of the algebraic aspects, including Hopf algebra cohomology, see
\cite{fauser:jarvis:2006a}.

\subsection{Multiplicativity of the coproducts}

A remarkable fact is the following
\mybenv{Proposition}
The coproduct of multiplication $\DM$ is a multiplicative function.
\myeenv

\noindent
{\bf Proof:} First consider relative prime numbers $p^r$, $q^s$
\begin{align}
\DM(p^r\cdot q^s)
&=\sum_{d\mid p^r\cdot q^s} d\times \frac{p^r\cdot q^s}{d}
\end{align}
since $p^r\mid q^s=1$, from which follows $d\mid p^r\cdot q^s=a\mid p^r\cdot 
b\mid q^s$, we obtain
\begin{align}
\DM(p^r\cdot q^s)
&=\sum_{{a\mid p^r}\atop{ b\mid q^s}}a\cdot b \times 
  \frac{p^r\cdot q^s}{a\cdot b} 
 = \sum_{l=0}^r \sum_{k=0}^s p^l q^k \times p^{r-l}q^{s-k}\nn
&=\sum_{l=0}^r p^l\times p^{r-l}\sum_{k=0}^s q^k\times q^{s-k}
 =\sum_{c\mid p^r} c\times\frac{p^r}{c}
  \sum_{d\mid q^s} d\times\frac{q^s}{d}\nn
&=\DM(p^r)\DM(q^s)
\end{align}
$\DM$ is not complete multiplicative due to
\begin{align}
\DM(4)&=1\times 4+2\times 2+4\times 1\nn
\DM(2)\DM(2)&=(1\times 2+2\times 1)^2
             =1\times 4+2\times 2+ 2\times 2+4\times 1
\end{align}
which completes the proof.
\qed

The lack of complete multiplicativity of the coproduct map spoils a major
axiom of Hopf algebra theory, namely the homomorphism axiom
\begin{align}\label{homAxiom}
\Delta\mu(n \times m) 
&= (\mu\otimes\mu)(\Id\otimes\mathsf{sw}\otimes\Id)
(\Delta\otimes\Delta)(n\times m)
\end{align}
which fails to hold(!) in the present case, but is only true as a 
multiplicative relation for $(n,m)=1$. The multiplicative convolution,
despite being bicommutative, biassociative, biunital, and having a nice
antipode, is alas not a Hopf algebra.

\section{Hopf gebra : multiplicativity versus complete multiplicativity}
\setcounter{equation}{0}

The fact that the convolution $\Conv(\cdot,\DM)$ is not a Hopf
algebra spoils the idea of employing a vast amount of standard machinery.
To distinguish the presently studied antipodal convolution
from a proper Hopf algebra we give it a new name.

\mybenv{Definition} A biassociative, biunital, antipodal convolution
$\Conv(\mu,\Delta,\antip)$ is called a Hopf gebra (HG). If the product is a
comultiplicative map and if the coproduct is a multiplicative map fulfilling
eqn. (\ref{homAxiom}) then the Hopf gebra is called 
multiplicative.\footnote{%
This notion is in the Bourbaki tradition \cite{Bourbaki:1989a} and was 
used in \cite{fauser:2002c} but originally coined by Oziewicz 
\cite{oziewicz:1997a}, however, with a different connotation.}
\myeenv

\subsection{Plan A: the modified crossing}

To be able to deal with the multiplicative, or even the general case, one
has to introduce new technical devices. For definitions etc. see
\cite{fauser:oziewicz:2001a,fauser:2002c}. A first attempt at a cure would
be to ask, if there could be a deformed crossing or switch
$c_{V,U} : V\otimes U\rightarrow U\otimes V$ so that the homomorphism axiom
eqn. (\ref{homAxiom}) could be reestablished in a complete multiplicative
fashion. This hope is nourished by the following

\mybenv{Theorem}[Oziewicz 97,\cite{oziewicz:1997a}] Every biassociative
antipodal convolution has a unique crossing $C_{V,U}$, such that $\Delta$
is a monoid homomorphism and $\mu$ is a comonoid homomorphism
\begin{align}
c_{V,U}=\quad
\pspicture[0.5](0,0)(3,4)
\psset{linewidth=\pstlw,xunit=0.5,yunit=0.5,runit=0.5,linecolor=black}
\psline{-}(1,8)(1,7)
\psline{-}(5,8)(5,7)
\psline{-}(1,1)(1,0)
\psline{-}(5,1)(5,0)
\psline{-}(0,6)(0,2)
\psline{-}(6,6)(6,2)
\psline{-}(3,5)(3,3)
\psarc(1,6){1.0}{0}{180}
\psarc(5,6){1.0}{0}{180}
\psarc(3,6){1.0}{180}{360}
\psarc(3,2){1.0}{0}{180}
\psarc(1,2){1.0}{180}{360}
\psarc(5,2){1.0}{180}{360}
\pscircle[linewidth=0.4pt,fillstyle=solid,fillcolor=white](1,7){0.2}
\pscircle[linewidth=0.4pt,fillstyle=solid,fillcolor=white](5,7){0.2}
\pscircle[linewidth=0.4pt,fillstyle=solid,fillcolor=white](3,5){0.2}
\pscircle[linewidth=0.4pt,fillstyle=solid,fillcolor=white](3,3){0.2}
\pscircle[linewidth=0.4pt,fillstyle=solid,fillcolor=white](1,1){0.2}
\pscircle[linewidth=0.4pt,fillstyle=solid,fillcolor=white](5,1){0.2}
\pscircle[linewidth=0.4pt,fillstyle=solid,fillcolor=white](0,4){0.2}
\pscircle[linewidth=0.4pt,fillstyle=solid,fillcolor=white](6,4){0.2}
\rput(-0.75,4){\antip}
\rput(5.25,4){\antip}
\endpspicture
\end{align}
If $c_{V,U}$ is a braid, i.e. 
$(c_{V,W}\otimes \Id_U)(\Id_V\otimes c_{U,W})(c_{U,V}\otimes \Id_W)
=(\Id_W\otimes c_{U,V})(c_{U,W}\otimes\Id_V)(\Id_U\otimes c_{V,W})$, 
om $U\otimes V\otimes W$, then the Hopf gebra is a braided Hopf algebra.
If $c_{U,V}$ is a (graded) switch the Hopf gebra becomes a (graded)
Hopf algebra.
\myeenv

The further route of such studies involves the possible classifications of
crossings obtained this way, and to detect if they are braided, compute their
minimal polynomial and so on. Such research is quite tedious, as was shown in
\cite{fauser:oziewicz:2001a}. The difficulties are so large, that in fact 
plan A has to be disregarded.

\subsection{Plan B: unrenormalization}

We need to come up with a new strategy. The idea is to reestablish a Hopf 
algebra structure as close as possible to the given multiplicative Hopf 
gebra in question. Then use the nicely behaved Hopf algebra for computations,
and try to find a transformation back to the Hopf gebra formulation. That
there is actually hope to do so, stems from the fact that we are going to
establish a Hopf algebra which is isomorphic to the multiplicative Hopf
gebra on all relatively prime inputs and differs only on common `overlapping'
prime factors. To comply with the usage of the term `renormalization' in
physics, we need to call such a map assigning to a multiplicative Hopf gebra
a Hopf algebra an `unrenormalization' map.

\mybenv{Definition} The unrenormalized coproduct of multiplication
$\DMU$ related to the (renormalized) coproduct of multiplication $\DM$
is recursively defined as
\begin{align} 
i)&& \DMU(p)&=\DM(p)=p\times 1+1\times p\quad\textrm{on primes}\nn
ii)&&\DMU(n\cdot m)&=\DMU(n)\cdot \DMU(m)\quad\forall n,m
\end{align}
forcing complete multiplicativity.
\myeenv
In this way the homomorphism axiom (\ref{homAxiom}) holds automatically
on all pairs $n,m$ of (non negative) integers. It is important to note
that this is a minimal alteration of the coproduct in the sense that the
unrenormalized coproduct differs only on the diagonal (on the $\gcd$'s)
from the original coproduct. While the counit still remains as the counit
of the unrenormalized coproduct, unrenormalization has, however, serious 
impacts, for example the
\mybenv{Corollary}
The unrenormalized antipode is given as $\SMU(n)=(-)^{\sum r_i}n$ where 
$n=\prod_i p_i^{r_i}$
\myeenv
This result shows that the antipode is just the grade involution with respect
to the grading of the natural numbers by prime number content. This is a
natural map in Hopf algebra theory, but far from being an interesting number
theoretic arithmetic function, like the M\"obius function, which was related to
the unrenormalized antipode.

We can now wonder, which duality connects multiplication and the new
unrenormalized coproduct.
\mybenv{Corollary}
Let $n=\prod_i p_i^{r_i}$ and $m=\prod_j p_j^{s_j}$. The pairing 
$(\ \mid\ )$ definjed by
\begin{align}
(n\mid m) &= \prod_i \delta_{r_i,s_i}r_i! \,=\, z_n
\end{align}
dualizes the multiplication $\cdot$ into $\DMU$.
\myeenv

\noindent
{\bf Proof:} (Sketch) Use Laplace expansion demanding that $\cdot$ and 
$\DMU$ are Milnor-Moore dual w.r.t. $(~\mid~)$. For details see
\cite{fauser:jarvis:2006a}.
\qed

Note that also in this case all alterations are just scalings:
$(n\mid m) = z_n \la n\mid m\ra$, which is up to a rescaling by
$z_n$ the Kronecker delta again. 

\subsection{The co-ring structure}

Before we try to set up the number theoretical model of renormalization, we
want to exhibit the co-ring structure. This implies a relation between the
coproduct of addition and the coproduct of multiplication in analogy with a
ring structure. Such relations were used in 
\cite{fauser:jarvis:king:wybourne:2005a} to investigate new group branchings.

Let us introduce a further group like coproduct $\delta : \openP \rightarrow
\openP\times \openP$, $\delta(p)=p\times p$. Coaddition and comultiplication 
are related as ($n=\prod_i p_i^{r_i}$)
\begin{align}
\DM(n) &= \sum_{d\vert n} d\times \frac{n}{d} \nn
       &= \delta^{\DP}(n) = \prod_i (p_i\times p_i)^{\DP(r_i)} \nn
       &= \sum_{r_i^\prime+r_i^{\pprime}=r_i}
          (\prod_i p_i^{r_i^\prime}) \times (\prod_i p_i^{r_i^{\pprime}})  
\end{align}
where the notion $\DM=\delta^{\DP}$ should be taken as a mnemonic only. 

The unrenormalized case follows along the same lines, and actually can be used
to define the unrenormalized coproduct of addition:
\begin{align}
\DPU(n)
&:=\sum_{n_1+n_2=n}\left( {n\atop n_1} \right) n_1 \times n_2
\end{align}
Let $n=\prod_i p_i^{r_i}$, unrenormalized addition and unrenormalized 
multiplication relate as:
\begin{align}
\DMU(n)
&=\delta^{\DPU}(n) \nn
&=\sum_{r_i^\prime+r_i^\pprime=r_i} 
\left( {r_i^\prime+r_i^\pprime\atop r_i}\right)
\prod_i p_i^{r_i^\prime}\times \prod_i p_i^{r_i^\pprime}
\end{align}
The appearance of the binomial factors is well known from calculations
in quantum field theories, describing the coproduct of for example scalar
fields.

\subsection{Coping with overcounting : renormalization}

Our paradigm is, that the number theoretically interesting structure is the
renormalized one, which is only multiplicative, and hence forms a 
multiplicative Hopf gebra (HG) only. To use nice algebraic machinery, we
associate to it an unrenormalized Hopf algebra (HA) which differs only on
common prime content, hence in a minimal way. The relation of the HG and
HA can be summarized in the following commutative diagram:
\vskip 2ex
\begin{align}\label{renCD}
~\hskip-0.75cm
\begin{array}{c@{\hskip 7truecm}c}
\psset{linewidth=\pstlw,xunit=0.25,yunit=0.25,runit=0.25,linecolor=white}
\Rnode{1}{HG(\cdot,\DM)} & \Rnode{2}{HA(\cdot,\DMU)} \\[10ex]
\Rnode{3}{HG(\cdot,\DM)} & \Rnode{4}{HA(\cdot,\DMU)}
\end{array}
\ncline[linecolor=black]{->}{1}{2}
\Aput{\text{unrenormalization}}
\ncline[linecolor=black]{->}{2}{4}
\Aput{\text{alg. manip. /pQFT}}
\ncline[linecolor=black]{->}{1}{3}
\Aput{\text{diff. comp. /NT}}
\ncline[linecolor=black]{->}{4}{3}
\Aput{\text{renormalization}}
\end{align}
\vskip 2ex
Number theoretical (NT) computations in the Dirichlet ring of arithmetic
functions, i.e. in the convolution ring over the Hopf gebra (HG), are 
performed along the left down arrow and are usually involved and complex.
Perturbative quantum field theory (pQFT) \textit{starts} with a Hopf algebra
structure assuming an algebra structure on the duals of the fields, either
explicitly or implicitly. Then algebraic calculations are performed
explicitly or implicitly using the underlying Hopf algebra (HA) structure.
However, the final formal expressions are plagued by infinities, which are
removed by a rescaling technique called renormalization. Our point is, that this
rescaling ends up in a Hopf gebra in analogy to number theory. The first step,
the unrenormalization is not seen in physics, since the modelling is done by
\textit{assuming} a Hopf algebra structure or equivalently a compatible 
algebra structure of the fields and their duals, which vice versa implies a
comultiplication. The technique of renormalization hence copes with
overcounting on the diagonals ($\gcd$ generalized to common maximal ideals).
In pQFT these overcountings are infinite, since summations are replaced
by integrations which in general diverge. In number theory one obtains finite
overcountings, and a Hopf algebra approach would have just failed to work
by producing wrong results. However, after having established the relation
of the diagram eqn. (\ref{renCD}) one is attempted to try to unrenormalize
problems in number theory and to use methods from QFT to handle them and
`renormalize' the formal result. Our approach opens at least two new
possibilities:

{\bf a:} pQFT starts with a HA structure, the unrenormalization
is hence superfluous. Due to scalings by counter terms
renormalization takes care of `overcounting the diagonal'. An enlargement
of modelling to start with unrenormalized quantities would possibly allow to
introduce number theoretic machinery, i.e. celebrated theorems and 
particular techniques, to solve problems in physics. 

{\bf b:} Via the unrenormalization, there may arise new possibilities
to deal which hard number theoretic problems in the ring of arithmetic
functions, by using methods from quantum field theory.

Hence renormalization should be understood as a sort of rewriting rule,
allowing insights to be moved from one side to the other.

There are several approaches to the theory of renormalization, discussed for
example in the topical review \cite{ebrahimi-fard:kreimer:2005a}. However,
from our point of view, the approach proposed by Brouder-Schmitt
\cite{brouder:schmitt:2002a}, based on Epstein-Glaser renormalization
\cite{epstein:glaser:1973a}, seems to be more natural and we have adopted
it in our work \cite{fauser:jarvis:2006a}. Therein it was shown for the
example of occupation number representations, that the ordering process
which we introduced in \cite{fauser:2001b} also applies for QM and used
both algebraic structures, the unrenormalized and renormalized ones.
Since the same process of deformation, but on another level of complexity,
produces the renormalization mechanism, we argue that the `ordering' or
`deformation' if done on the higher level of complexity --multiplication
versus addition, or composition versus multiplication-- enters at least in a
twofold manner, the more complex one giving rise to the renormalization map.
In terms of symmetric functions this leads to the Hopf algebra of plethysm
\cite{fauser:jarvis:2006b}.
The crucial fact is, that addition can be obtained as iteration of the successor
map, multiplication as the iteration of the addition and exponentiation as the
iteration of multiplication. Further generalization fails, since the 
iteration functor needs a transposition, which is equivalent to demanding
a commutative binary underlying operation \cite{lawvere:rosebrugh:2003a}.
In that sense, our number theoretic model needs to be enlarged to include 
exponentiation to actually parallel the `renormalization' encountered in
pQFT. 

\section{Taming multiplicativity}
\setcounter{equation}{0}

A complete multiplicative function $g$ is defined if its values are known 
on all primes, i.e. on $\openP$. Let $n=\prod_i p_i^{r_i}$ then complete
multiplicativity allows to write
\begin{align}
g(n)&=g(\prod_i p_i^{r_i})=\prod_i g(p_i)^{r_i}.
\end{align}
However, a multiplicative function $f$ needs to be specified on all prime
powers $\{p_i^k\}$, $\forall i,k$
\begin{align}
f(n)&=f(\prod_i p_i^{r_i})=\prod_i f(p_i^{r_i}).
\end{align}
While both sets have the same cardinality it is awkward that a multiplicative
function is not well defined by its values on generators, here the primes in
$\openP$.
  
In what follows, we want show how one might tame multiplicativity by giving
data \textit{only} on primes, and controlling the values on $f(p^n)$ by a
recursion involving a complete multiplicative function. This idea is based
on the analogy, that the expectation values of powers of quantum fields 
$\la 0\mid (\psi(x))^{2}\mid 0\ra$ should be computable from a function
of the expectation values $\la 0\mid \psi(x)\psi(y)\mid 0\ra$ in a suitable
limit $y\rightarrow x$.

The device we want to use is that of Bell series. These are series encoding
an arithmetic function on all prime powers of a given prime $p$.
\mybenv{Definition}
A Bell series of an arithmetic function $f$ for a fixed prime $p$ is
given as an ordinary power series
\begin{align}
f_p(x) =\sum_{n\ge0} f(p^n)x^n
\end{align}
employing a formal indeterminate $x$.
\myeenv

\mybenv{Corollary} If $f$ is complete multiplicative its Bell series
reads
\begin{align}
f_p(x)&= \sum f(p)^n\,x^n=\frac{1}{1+f(p)x}
\end{align}
\myeenv
\noindent
The Bell series of the M\"obius function and the Euler totient function 
read
\begin{align}
\mu_p(x)&=1-x \nn
\phi_p(x)&=\frac{1-x}{1-px}
\end{align}
showing that they are not complete multiplicative. The most important fact about
Bell series for us is, that the Dirichlet convolution product of arithmetic
functions is transformed into the Cauchy product of Bell series.
Let $h=f\conv g$, then 
\begin{align}
h_p(x)=f_p(x)g_p(x)
\end{align}
reducing the complexity of the operation dramatically. 

We use an example from Apostol \cite{apostol:1979a} to demonstrate how this
might be used to model the process of renormalization in number theory in
analogy to renormalization in pQFT, by adding counter terms or modifying
the pole structure of the `propagator'.

Let $g$ be complete multiplicative, recall that then $g(1)=1$. We define a 
recursion for a multiplicative function $f$ so that all values of $f$ on 
prime powers are determined. In terms of coefficients a particular recursion
reads 
\begin{align}\label{a}
f(p^{n+1}) &= f(p)f(p^n)-g(p)f(p^{n-1}).
\end{align}
This allows to compute the Bell series
\begin{align}\label{b} 
f_p(x) = \frac{1}{1-f(p)x+g(p)x^2}.
\end{align}
It can be shown, that eqn. (\ref {b}) follows from eqn. (\ref{a}) and vice
verca. Using this recursion, it is now possible to establish the following
product formula
\begin{align}\label{c}
f(m\cdot n) 
&= f(m)f(n)
 - \sum_{{d\mid \textsf{gcd}(n,m)}\atop{1<d\le\textsf{gcd}(n,m)}} 
         g(d)f(\frac{m\cdot n}{d^2})
\end{align}
Together with eqn. (\ref{b}) this establishes a number theoretic analog 
of renormalization theory. $g(p)$ would serve as an additive renormalization
of the `propagator' $\frac{1}{1-f(p)x+g(p)x^2}$ and the sum in the
right hand side of eqn. (\ref{c}) constitutes counter terms.

\noindent\textbf{Acknowledgements.} It is a pleasure to thank Peter Jarvis
for many helpful discussions and for ongoing collaboration on this subject.
Part of this work was done in Hobart during a visit supported by the ARC
research grant DP0208808, and the Alexander von Humboldt Foundation.

\begin{appendix}
\section{Some facts about Dirichlet and Bell series}

\subsection{Characterizations of complete multiplicativity}

Since multiplicativity versus complete multiplicativity plays a major role in
our argumentation we want to recall useful characterizations of
multiplicativity. 

Lambek \cite{lambek:1966a} proved that an arithmetical function $f$ is
completely multiplicative iff its Hadamard product distributes over every
Dirichlet product:
\begin{align}
f.(g\conv h)
&= (f.g)\conv (f.h)
\end{align}
for all arithmetical functions $g,h$. In terms of coefficients this reads
\begin{align}
f(n)\sum_{d\mid n} g(d)h(n/d) 
&= \sum_{d\mid n} f(d)g(d) f(n/d)h(n/d)
\end{align}
Carlitz \cite{carlitz:1971a} posed the problem to characterize complete 
multiplicativity by distributivity over particular Dirichlet convolutions.
Let $\tau=\zeta\conv\zeta$ be the number of positive divisors function. $f$ 
is complete multiplicative iff
\begin{align}
f.\tau 
&= (f.\zeta)\conv(f.\zeta) 
 = f\conv f
\end{align}
A nice way to generalize such notions is by using M\"obius categories
$\mathcal{C}$ \cite{leroux:1975a,leroux:1990a}. 
These are categories defined to generalize and unify the theory of M\"obius
inversions. In terms of morphisms one investigates incidence functions 
$f,g$ forming an incidence algebra $A(\mathcal{C})$ by employing the product
\begin{align}
(f\conv g)(\alpha)
&= \sum_{\alpha^\prime\alpha^{\pprime}=\alpha}
   f(\alpha^\prime)g(\alpha^{\pprime})
\end{align}
An incidence function is complete multiplicative iff 
\begin{align}
f(\alpha) 
&= f(\alpha^\prime)f(\alpha^{\pprime})
\end{align}
with $\alpha=\alpha^\prime\alpha^{\pprime}$ the composition of morphisms. Now, 
Lambek's characterization generalizes to this case, while Carlitz' 
characterization has to be altered \cite{schwab:2004a}. It is, however, nice
to have a generalization to this general setting allowing to export the
concept of multiplicativity to incidence (or functionals on operator)
algebras. This way of generalization is needed on the way to establish 
our analogy between number theory and pQFT in a more concrete way.

\subsection{Groups and subgroups of Dirichlet convolution}

This section follows the exposition of Dehaye \cite{dehaye:2002a}.
Let $\mathsf{F}_0$ be the set of multiplicative functions different from
the zero function $\mathbf{0}(n)=0$ for all $n$. This amounts to have 
$f(1)=1$ for all $f$ in $\mathsf{F}_0$. The pair $(\mathsf{F}_0,\conv)$
is an abelian group with Dirichlet convolution as product.

For any prime $p$ we define $\mathsf{F}^p=\{ f\in \mathsf{F}_0 \mid
f(n)=0 \textrm{~~for every~~$n$~~s.t.~~} p\kern-1.0ex\not{\kern-0.5ex\vert}
\,n\}$. That is an $f\in \mathsf {F}^p$ has support on prime powers $p^k$
only. For every prime $p$, $(\mathsf{F}^p,\conv)$ is a subgroup of
$\mathsf{F}_0$. Furthermore, there exists an isomorphism between
$\mathsf{F}^p$ and the group of upper-triangular non-zero infinite
matrices $\mathbf{M}^1$
\begin{align}
\mathbf{M}^1
&= \{ m\in \mathbf{M} \mid m(a,a)=1, \forall a\in \openN, \text{~and~}
   m(a,b)=0, \forall a,b\in \openN \text{~s.t.~} a>b \} 
\end{align} 
$\mathbf{M}^1$ is a group with the infinite unit matrix as identity element. 
The isomorphism $\phi : \mathsf{F}^p \rightarrow \mathbf{M}^1$ is given by
\begin{align}
\phi(f) =\left(\begin{array}{ccccc}
1 & f(p) & f(p^2) & f(p^3) & \cdots \\
0 &    1 & f(p)   & f(p^2) & \cdots \\
0 &    0 &      1 & f(p)   & \cdots \\
0 &    0 &      0 & 1      & \cdots \\
\vdots & \vdots & \vdots & \vdots & \ddots
\end{array}\right)
\end{align}
and
\begin{align}
\phi(f)\cdot \phi(g) &= \phi(f\conv g)
\end{align}
where in the l.h.s. the product is matrix multiplication. This isomorphism
shows that the Bell series are particular Dirichlet series or restrictions of
Dirichlet series to the subgroup $\mathsf{F}^p$. These groups are isomorphic
for every pair of primes $p_i,p_j$.

It is possible to consider $\mathsf{F}_0$ as a complete (or Cartesian) direct
product of the subgroups $\mathsf{F}^{p_i}$ for all primes
\begin{align}
\mathsf{F}_0 &= \overline{\prod}_{i\in\openN} \mathsf{F}^{p_i}
\end{align}
It can be shown \cite{dehaye:2002a} that
\begin{itemize}
\item[a)] The group $\mathsf{F}_0$ is torsion free (i.e. has no element of
finite order).
\item[b)] $\mathsf{F}^p$ is a subgroup of $\mathsf{F}_0$ for every prime $p$,
all such subgroups are pairwise isomorphic and are isomorphic to infinite
upper-triangular non-zero matrices or to Bell series.
\item[c)] $\mathsf{F}_0$ is isomorphic to the complete direct product of the
subgroups $\mathsf{F}^p$.
\item[d)] $\mathsf{F}_0$ is divisible and has a natural structure of a vector
space over $\openQ$.
\end{itemize}

\section{\label{Petermann}Densities of generators}

We want to emphasise another point which connects our work with renormalization
of quantum fields. As our discussion here and in \cite{fauser:jarvis:2006a}
demonstrated, one can grade the natural numbers in two canonical ways attached
to addition and multiplication. If we generate the natural numbers additively,
we have but one generator, the one $1$, which is the target of the successor
map, and all numbers are generated as successors of the zero $0$. The successor
map is assumed to have no torsion and composition is associative. From this
construction it is evident, that the density of natural numbers in the natural
numbers $d(n,n_0)$ with respect to the $1$ as generator is constant. That is
in every neighbourhood, that is an interval containing $n_0$, of a natural
number $n_0$ one finds the same density of natural numbers.

The second way to grade the natural numbers was induced by the multiplicative
structure and the primitive elements, i.e. generators, were shown to be the set
of prime numbers $\{p_i\}$. However, the density of prime numbers in the
natural numbers is a nontrivial function. The celebrated prime number theorem
states that the number of prime numbers below $n_0$ is $\frac{n_0}{\log n_0}$
for $n_0\rightarrow \infty$. This renders it obvious that a multiplicative
construction of the integers behave quite differently with respect to the
densities of generators in the natural numbers. A.~Petermann showed in a
remarkable paper \cite{petermann:2000a} that a renormalization group analysis
provides a proof for the prime number theorem. This supports our claim, that
the present simplified model of renormalization is actually rich enough to
contain main features of renormalization in quantum field theory.

It is possible to iterate this process by asking what kind of `primitive
elements' occur if one looks for exponentiation as an iteration of
multiplication. This question leads into the realm of modular forms and one
obtains higher order corrections in the densities of `higher primitive
elements' along the same lines as one obtains higher order loop corrections,
and divergencies in perturbative quantum field theory. This will be explored
elsewhere. 
\end{appendix}

{\small

\begin{thebibliography}{10}

\bibitem{apostol:1979a}
Tom~M. Apostol.
\newblock {\em {Introduction to Analytic Number Theory}}.
\newblock Springer-Verlag, New York, 1979.
\newblock [fouth printing 1995].

\bibitem{Bourbaki:1989a}
Nicolas Bourbaki.
\newblock {\em Elements of {M}athematics: {A}lgebra I -- {C}hapters 1--3}.
\newblock Springer-Verlag, Berlin, 1989.

\bibitem{brouder:schmitt:2002a}
Christian Brouder and William Schmitt.
\newblock {Quantum groups and quantum field theory III. Renormalization}.
\newblock {\em preprint}, pages 1--18, 2002.
\newblock hep-th/0210097.

\bibitem{bruedern:1995a}
J\"org Br\"udern.
\newblock {\em {Einf\"uhrung in die analytische Zahlentheorie}}.
\newblock Springer-Verlag, Berlin, 1995.

\bibitem{carlitz:1971a}
L.~Carlitz.
\newblock {Problem E 2268}.
\newblock {\em Amer. Math. Monthly}, 78:1140, 1971.

\bibitem{dehaye:2002a}
Paul-Olivier Dehaye.
\newblock {On the structure of the group of multiplicative arithmetical
  functions}.
\newblock {\em Bull. Belg. Math. Soc. Simon Stevin}, 9(1):15--21, 2002.

\bibitem{ebrahimi-fard:kreimer:2005a}
Kurusch Ebrahimi-Fard and Dirk Kreimer.
\newblock {The Hopf algebra approach to Feynman diagram calculations}.
\newblock {\em J. Phys. A: Math. Gen.}, 38:R385--R407, 2005.
\newblock Topical Review.

\bibitem{epstein:glaser:1973a}
H.~Epstein and V.~Glaser.
\newblock The role of locality in perturbation theory.
\newblock {\em Ann. Inst. Henri Poincar\'e}, 19:211--295, 1973.

\bibitem{fauser:2001b}
Bertfried Fauser.
\newblock {On the Hopf-algebraic origin of Wick normal-ordering}.
\newblock {\em Journal of Physics A: Mathematical and General}, 34:105--115,
  2001.
\newblock hep-th/0007032.

\bibitem{fauser:2002c}
Bertfried Fauser.
\newblock {A Treatise on Quantum Clifford Algebras}.
\newblock Konstanz, 2002.
\newblock Habilitationsschrift, arXiv:math.QA/0202059.

\bibitem{fauser:jarvis:2006a}
Bertfried Fauser and P.D. Jarvis.
\newblock {The Dirichlet Hopf algebra of arithmetics}.
\newblock {\em Journal of Knot Theory and its Ramificatiuons}, pages 1--42,
  2006.
\newblock accepted; math-ph/0511079.

\bibitem{fauser:jarvis:2006b}
Bertfried Fauser and P.D. Jarvis.
\newblock {The Hopf algebra of plethysms}.
\newblock {\em work in progress}, 2006.

\bibitem{fauser:jarvis:king:wybourne:2005a}
Bertfried Fauser, Peter~D. Jarvis, Ron~C. King, and Brian~G. Wybourne.
\newblock {New branching rules induced by plethysm}.
\newblock {\em J. Phys A: Math. Gen.}, pages 1--40, 2006.
\newblock accepted; math-ph/0505037.

\bibitem{fauser:oziewicz:2001a}
Bertfried Fauser and Zbigniew Oziewicz.
\newblock {Clifford Hopf gebra for two dimensional space}.
\newblock {\em Miscellanea Algebraicae}, 2(1):31--42, 2001.
\newblock math.QA/0011263.

\bibitem{lambek:1966a}
J.~Lambek.
\newblock {Arithmetical functions and distributivity}.
\newblock {\em Amer. Math. Monthly}, 73:969--973, 1966.

\bibitem{lawvere:rosebrugh:2003a}
F.~William Lawvere and Robert Rosebrugh.
\newblock {\em {Sets for mathematics}}.
\newblock Cambridge Univ. Press, Cambridge, 2003.

\bibitem{leroux:1975a}
P.~Leroux.
\newblock {Les Cat\'egories M\"obius}.
\newblock {\em Cahiers Topologie G\'eom. Diff\'erentielle Cat\'eg.},
  16:280--282, 1975.

\bibitem{leroux:1990a}
P.~Leroux.
\newblock {Reduced matrices and $q$-$\log$-concavity properties of $q$-Stirling
  numbers}.
\newblock {\em J. Combin. Theory Ser. A}, pages 64--84, 1990.

\bibitem{oziewicz:1997a}
Zbigniew Oziewicz.
\newblock Clifford {H}opf gebra and biuniversal {H}opf gebra.
\newblock {\em Czechoslovak Journal of Physics}, 47(12):1267--1274, 1997.
\newblock q-alg/9709016.

\bibitem{petermann:2000a}
A. Petermann.
\newblock The so-called Renormalization Group method applied to the specific
  prime numbers logarithmic decrease
\newblock {\em Eur. Phys. J.} C 17:367--369, 2000

\bibitem{schwab:2004a}
Emil~Daniel Schwab.
\newblock {Characterizations of Lambek-Carlitz type}.
\newblock {\em Archivum Mathematicum (Brno)}, 40:295--300, 2004.

\bibitem{selberg:1949a}
A.~Selberg.
\newblock {An elementary proof of the prime number theorem}.
\newblock {\em Ann. Math.}, 50:305--313, 1949.

\end{thebibliography}

}

\end{document}